\documentclass[aps,prb,reprint,amsmath,amssymb,showpacs,superscriptaddress,twocolumn]{revtex4}
\usepackage{graphicx}
\usepackage{xcolor}
\usepackage{subfigure}
\usepackage{hyperref,hypcap}
\usepackage{braket}
\usepackage{amsmath}

\begin{document}
\title{Microscopic Theory of Equilibrium Polariton Condensates}
\author{Fei Xue}
\author{Fengcheng Wu}
\author{Ming Xie} 
\affiliation{Department of Physics, The University of Texas at Austin, Austin, TX 78712, USA}
\author{Jung-Jung Su} 
\affiliation{Department of Electrophysics, National Chiao Tung University, Hsinchu 300, Taiwan}
\author{A.H. MacDonald}
\affiliation{Department of Physics, The University of Texas at Austin, Austin, TX 78712, USA}

\date{\today}

\begin{abstract}
We present a microscopic theory of the equilibrium polariton condensate state 
of a semiconductor quantum well in a planar optical cavity.  The theory accounts for the
adjustment of matter excitations to the presence of a coherent photon field,
predicts effective polariton-polariton interaction strengths that are weaker and condensate exciton fractions that 
are smaller than in the commonly employed exciton-photon model, and yields  
effective Rabi coupling strengths that depend on the detuning of the cavity photon energy 
relative to the bare exciton energy.  The dressed quasiparticle bands that appear 
naturally in the theory provide a mechanism for electrical manipulation of 
polariton condensates.  
\end{abstract}

\pacs{71.35.-y, 73.21.-b, 71.36.+c, 71.35.Lk}
\maketitle

\section{Introduction}

\begin{figure}[t]
	\includegraphics[width=1\columnwidth]{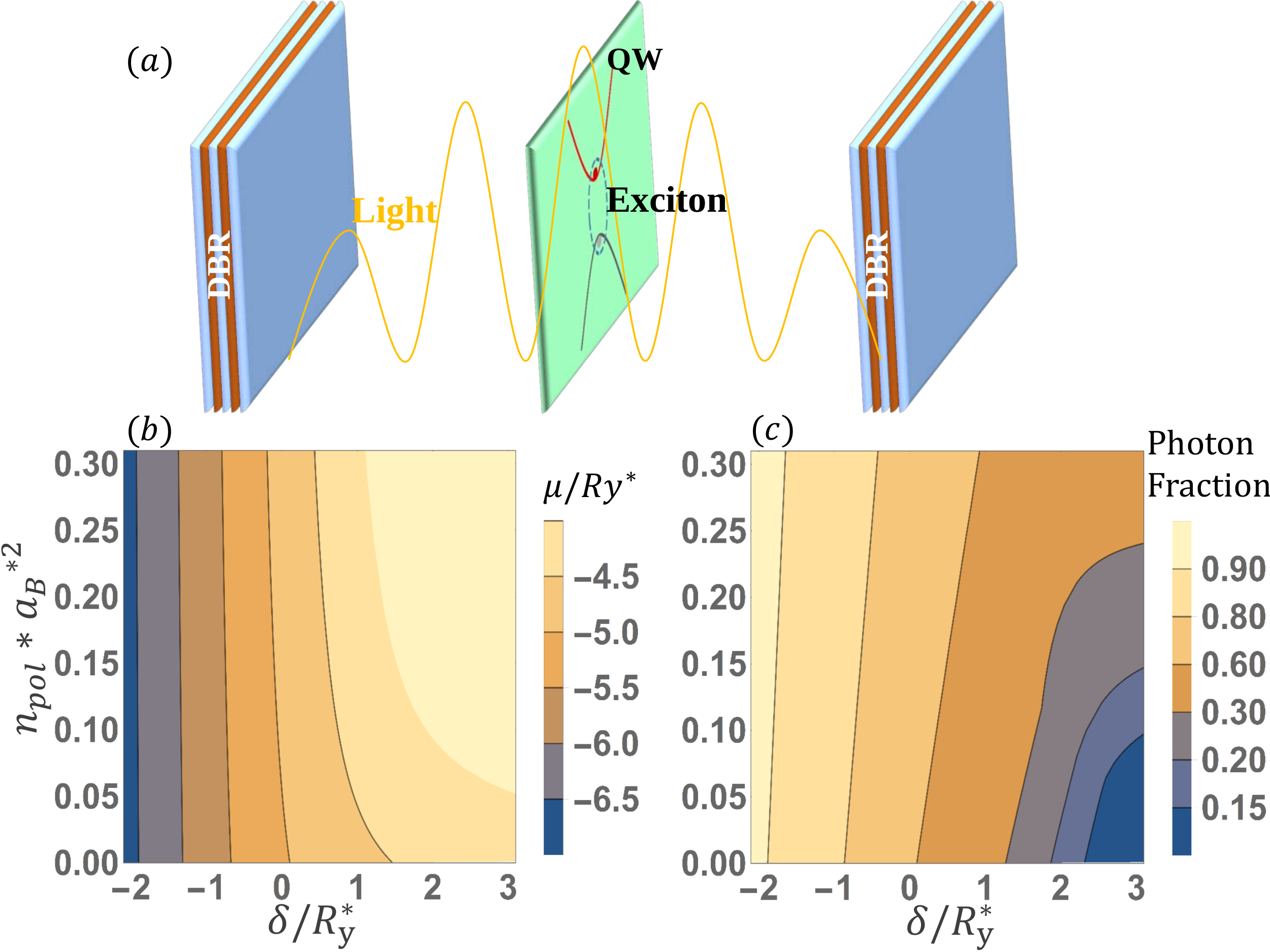}
	\caption{(Color online) 
	(a): Typical polariton condensate geometry.  One or several
	quantum wells are placed between a pair of distributed Bragg reflectors (DBRs). 
	The two-dimensional quantum well conduction and valence band states are dressed by their interactions with 
	condensed cavity photons and by electron-electron interactions.
	(b) and (c): Polariton condensate chemical potential and photon fraction as a function of detuning $\delta$
		and polariton density $n_{pol}$.
	}
	\label{Fig:PhaseDiagram}
\end{figure}

A polariton is a quantum state in which a photon is coherently mixed with an elementary 
excitation of condensed matter, for example an exciton in a semiconductor or a surface plasmon
in a metal.  Two-dimensional polariton condensate states
can be formed\cite{Tassone1999,Deng2002,Malpuech2003,Kasprzak2006,Snoke2007,Keeling2007,Amo2009,Byrnes2010, Deng2010,Kamide2010,Kamide2011,Byrnes2014}
when semiconductor quantum wells are placed in a 
planar optical cavity\cite{Skolnick1998} and pumped to create populations of $\vec{q}=0$ cavity photons 
and $\vec{q}=0$ quantum well excitations.  (See Fig.~\ref{Fig:PhaseDiagram}.)
In the condensed state the photon and 
quantum well excitation states are both separately and mutually coherent.
When the scattering rates between states formed by 
the quantum well excitations and the cavity-photons exceed\cite{Deng2006,Littlewood2006,Wouters2007,Byrnes2014}  
polariton lifetimes, a circumstance that is regularly achieved,\cite{Deng2002,Kasprzak2006,Snoke2007,Snoke2015} 
the polariton condensate steady state can be described microscopically 
using equilibrium statistical mechanics.  In this paper we present a fully microscopic theory 
of equilibrium polariton condensates that treats the two-dimensional quantum well
band states explicitly and goes beyond the commonly used model in which 
bare excitons are treated as Bose particles that are coupled via flip-flop interactions with 
cavity photons.  We find that the effective polariton-polariton interaction strength is weaker and that the condensate 
exciton fraction is smaller than in the commonly employed exciton-photon theory of a polariton 
condensate, and that the quasiparticle bands are more strongly dressed for a given 
polariton density at positive detuning $\delta$ than at negative detuning.  
Similar calculations were preformed previously\cite{Kamide2010,Byrnes2010,Kamide2011,Yamaguchi2013}
with the goal of shedding light on the BEC-BCS crossover of exciton-polariton condensates.
In this paper, we are motivated by recent pioneering work on electrical coupling to polariton
condensates,\cite{Tartakovskii2015} anticipating that the 
polariton dressing of the quantum-well band states on which we focus 
provides a mechanism for electrical manipulation of polariton condensates.

Some of our principle results are summarized in Fig.~\ref{Fig:PhaseDiagram} (b) and (c) in which we 
plot the polariton chemical potential and the polariton photon-fraction 
as a function of detuning and polariton density.  We will compare these results, and 
others, with the predictions of the simplified bosonic exciton-photon theory.\cite{Eastham2001,Keeling2007}
Our paper is organized as follows. In Sec. \ref{theory} we explain our 
formulation of the microscopic equilibrium polariton condensate theory,
which differs somewhat from the one employed in previous work.  In Sec. \ref{results} we present and discus
results obtained for equilibrium polariton condensate properties 
using this approach, comparing where possible with the corresponding results 
implied by the simplified theory.  Finally in Sec. \ref{discussion} we present our 
conclusions and comment on potential applications of 
coherent electrical coupling to polariton condensates.

\section{Equilibrium Polariton Condensates} 
\label{theory}
For simplicity we consider a polariton condensate system with a 
single quantum well and nelgect the electronic spin degree-of-freedom.
The Hamiltonian of the quantum-well/cavity-photon system is then 
\begin{equation} 
\hat{H}_{QWCP}=\hat{H}_{mat}+\hat{H}_{ph} + \hat{H}_{mat-ph},
\end{equation} 
where
\begin{equation}
\label{Hamiltonian}
\begin{split}
&\hat{H}_{mat}=\sum_{\vec{k}} \, \bigg[ (E_c + \frac{\hbar^2k^2}{2m_e}) \, a_{c\vec{k}}^{\dagger} a_{c\vec{k}} +
(E_v -\frac{\hbar^2k^2}{2m_h} ) a_{v\vec{k}}^{\dagger} a_{v\vec{k}} \bigg] \\
&+\frac{1}{2A}\sum_{\vec{k},\vec{k'},\vec{q}} \, V_{\vec{q}} \,
\bigg[a_{c\vec{k}+\vec{q}}^{\dagger}a_{c\vec{k'}-\vec{q}}^{\dagger}a_{c\vec{k'}}a_{c\vec{k}}
+a_{v\vec{k}+\vec{q}}^{\dagger}a_{v\vec{k'}-\vec{q}}^{\dagger}a_{v\vec{k'}}a_{v\vec{k}} \\
&   - 2a_{c\vec{k}+\vec{q}}^{\dagger}a_{v\vec{k'}}a_{v\vec{k'}-\vec{q}}^{\dagger}a_{c\vec{k}} \bigg] , \\
& \hat{H}_{mat-ph} =   -\frac{g}{\sqrt{A}} \sum_{\vec{k},\vec{q}} (a_{c\vec{k}+\vec{q}}^{\dagger}a_{v\vec{k}} \Phi_{q} 
+ a_{v\vec{k}}^{\dagger} a_{c\vec{k}+\vec{q}} \Phi_{q}^{\dagger} ),\\
&\hat{H}_{ph}=\sum_{\vec{q}}  \Phi_{\vec{q}}^{\dagger}  \Phi_{\vec{q}}   (\epsilon_{ph} + \frac{\hbar^2q^2}{2m_{ph}}),
\end{split}
\end{equation}
$\Phi_{q}^{\dagger}$ and $\Phi_{q}$ are cavity photon creation and annihilation operators,
$\epsilon_{ph}$ is the $\vec{q}=0$ cavity photon energy, 
$a_{c,v\vec{k}}^{\dagger}$ and $a_{c,v\vec{k}}$ are quantum well conduction and valence 
band electron creation and annihilation operators, $m_{ph}$ is the cavity 
photon mass, $A$ is the two-dimensional system area,
and $V_{\vec{q}} = 2\pi e^2/ \epsilon q$ is the repulsive two-dimensional Coulomb interaction.

Because it neglects photon leakage from the optical cavity, and the weak purely-electronic or 
phonon-mediated disorder and interaction processes that can transfer electrons between conduction and valence bands,
the quantum-well/cavity-photon Hamiltonian conserves 
not only electron number but also the sum of the number of photons and the number of electrons that are promoted from the  
valence band to the conduction band, ({\it i.e.} the number of matter excitations).
We therefore define the number of polaritons as the sum of the number of matter excitations and
the number of photons: 
\begin{eqnarray} 
N_{pol} &=& N_{ph} + N_{ex} \nonumber \\ 
&=& \sum_{\vec{k}}  \bigg[ (a_{c\vec{k}}^{\dagger} a_{c\vec{k}} + a_{v\vec{k}} a_{v\vec{k}}^{\dagger})/2 + 
\Phi_{\vec{k}}^{\dagger}  \Phi_{\vec{k}}   \bigg], 
\end{eqnarray} 
and observe that both $[\hat{H}_{QWCP}, N]$ and $[\hat{H}_{QWCP}, N_{pol}]$ vanish.
Below we define $N$ as the total electron number relative to the number 
present in the neutral state with filled valence
bands and empty conduction bands. 

We now use mean-field theory to approximate the ground state of $\hat{H}_{QWCP}$ in the 
Fock space sector with $N=0$ and $N_{pol} = n_{pol} A$ equal to an extensive value proportional to the sample area $A$.  
The constraint on $N_{ex} + N_{ph} = A (n_{ex} + n_{ph}) $ is most conveniently enforced by first fixing the 
number of photons and then using an exciton chemical potential to enforce the constraint on 
$N_{ex}$.  

In our mean-field approximation all cavity photons in the equilibrium polariton condensate occupy the lowest energy
$\vec{q}=0$ state, and electron-electron interactions are approximated using Hartree-Fock theory. 
These approximations lead to the following mean-field Hamiltonian for the matter subsystem 
\begin{equation}
H_{MF}=\sum_{\vec{k}}
(a_{c\vec{k}}^{\dagger},a_{v\vec{k}}^\dagger)
(\zeta_{\vec{k}}+\xi_{\vec{k}} \sigma_z-\Delta_{\vec{k}} \sigma_x)
\begin{pmatrix} a_{c\vec{k}} \\ a_{v\vec{k}}\end{pmatrix}
\label{MF}
\end{equation}
where $\sigma_{z,x}$ are Pauli matrices that act on coherently mixed spinors with conduction
and valence band components and the dressed band parameters $\xi_{\vec{k}}$ and $\Delta_{\vec{k}}$, are 
obtained by solving the self-consistent-field equations:
\begin{equation}
\label{SC}
\begin{split}
&\xi_{\vec{k}}=\frac{\hbar^2k^2}{4m} + \frac{E_{gap} - \mu}{2}-\frac{1}{2A}\sum_{\vec{k}'}V_{\vec{k}-\vec{k}'}(1-\xi_{\vec{k}'}/E_{\vec{k}'}),\\
&\Delta_{\vec{k}}=\frac{1}{2A}\sum_{\vec{k}'}V_{\vec{k}-\vec{k}'}\frac{\Delta_{\vec{k}'}}{E_{\vec{k}'}}+g\sqrt{n_{ph}},\\
&E_{\vec{k}}=\sqrt{\xi_{\vec{k}}^2+\Delta_{\vec{k}}^2},
\end{split}
\end{equation}
where $m=m_em_h/(m_e+m_h)$ is the reduced mass, $n_{ph}$ is the density 
of photons, and $E_{gap}=E_c-E_v$ is the gap between conduction and valence band.  The terms in Eq.~\ref{SC} containing $V_{\vec{k}-\vec{k}'}$ factors
are electron-electron interaction self-energies.
(Note that the band energies in Eq.~\ref{Hamiltonian} are defined as the quasiparticle energies in the 
state with no electrons in the conduction band and no holes in the valence band.)
The term proportional to $\zeta_{k}=\hbar^2k^2[1/(4m_e)-1/(4m_h)]$
in Eq.~\ref{MF} accounts for the effective mass 
difference between conduction and valence bands and plays no role in the 
excitation spectrum because it simply adds a constant to the many-body energy at zero temperature. 

These mean-field equations are identical to those that appear in the theory of purely excitonic condensates, apart 
from the contribution $ g \sqrt{n_{ph}}$ to the self-energy $\Delta_{\vec{k}}$.
This term adds to electronic self-energies in supporting coherence between conduction and valence band states in the 
dressed quantum-well bands.\cite{Keldysh1965,Lozovik1976,Comte1982,Zhu1995}
As emphasized in earlier work,\cite{Kamide2010,Byrnes2010,Kamide2011,Yamaguchi2013} because the coupling to the 
photon field is independent of momentum in the $\vec{k} \cdot \vec{p}$ theory we use,  which is 
accurate for all systems of interest,  it yields electron-hole pairs that are more tightly bound than 
they would be if only electron-electron interactions were present.
In these equations we have already enforced the 
$N=0$ electron-number constraint by occupying only dressed valence band 
states in constructing the electron-electron interaction self-energies.  Below 
we will measure excitation energies relative to $E_{gap}$, thereby setting the zero for matter excitation 
energies at the quantum well energy gap.  

After solving Eq.(\ref{SC}) self-consistently, we can evaluate the  
exciton density $n_{ex}$ and the matter energy
per area $\epsilon_{mat}=(\braket{\hat{H}_{mat}}+\braket{\hat{H}_{mat-ph}})/A$ 
as a function of the exciton chemical potential $\mu$ and the density of photons $n_{ph}$:  
\begin{equation}
\label{nex}
n_{ex}=\frac{1}{2A}\sum_{\vec{k}}(1-\xi_k/E_k),
\end{equation} 
\begin{equation}
\label{emat}
\begin{split}
\epsilon_{mat}=&\frac{1}{2A}\sum_{\vec{k}}\bigg[(\frac{\hbar^2k^2}{4m}+\frac{\mu}{2}+\xi_k)(1-\frac{\xi_k}{E_k})\\
&-(g\sqrt{n_{ph}}+\Delta_k)\frac{\Delta_k}{E_k}\bigg].
\end{split}
\end{equation}
Note that all quantities are functions of wavevector magnitude $k$ only,
since the excitons condense in an $s$-wave state.

In a quasi-equilibrium polariton condensate light and matter 
share a common chemical potential $\mu$.
In order to enforce this mutual equilibrium between the photon and the quantum-well excitation 
parts of the condensate we need to evaluate the photon chemical potential and set it equal to the excitation chemical potential.
It follows that for a given $n_{ph}$ and $\mu$, 
\begin{equation}
\label{equilibrium}
\mu_{ph} =\frac{\partial \braket{\hat{H}_{QWCP}}}{\partial N_{ph}} = \epsilon_{ph} + \frac{\partial \epsilon_{mat}}{\partial n_{ph}} = \mu.
\end{equation}
We follow normal practice in expressing the cavity photon energy in terms 
of the detuning $\delta$, defined as the difference between $\epsilon_{ph}$ and 
the energy of a single isolated exciton $\epsilon_{ex}$.  With our choice of the quantum well
band gap as the zero of excitation energy 
$\epsilon_{ex}=-E_b$ and $\epsilon_{ph} = \delta - E_{b}$ where $E_{b}$ is the exciton binding energy.
In the illustrative calculations performed below,  which do not correct for the finite width of the quantum well, 
$E_{b} =  4 Ry^* = 2 \hbar^2 /m a_B^2$, where $Ry^*$ is the 
semiconductor Coulomb energy scale and $a_{B}^* = \hbar^2 \epsilon/(m e^2)$ is the corresponding 
length scale.  

For any given value of $n_{ph}$ and $\mu$, the quantum well excitations and the 
photons are in mutual equilibrium at some value of the detuning energy $\delta$.  
We therefore solve the matter equations self-consistently over a range of $n_{ph}$ and 
$\mu$ values and evaluate $\mu_{ph}$ by using a Hellman-Feynman expression for 
the derivative in Eq.~\ref{equilibrium},
\begin{equation}
\frac{\partial \epsilon_{mat}}{\partial n_{ph}} = \braket{\frac{\partial \hat{H}_{mat-ph}}{\partial \hat{N}_{ph}}} 
= -\frac{1}{A}\frac{g}{\sqrt{n_{ph}}}\sum_{\vec{k}} u_{\vec{k}} v_{\vec{k}}, 
\end{equation}
where $u_{\vec{k}}=\sqrt{\frac{1}{2}(1 + \xi_{\vec{k}}/E_{\vec{k}})}$ and 
$ v_{\vec{k}} =\sqrt{\frac{1}{2}(1 - \xi_{\vec{k}}/E_{\vec{k}})}$ are the bare valence 
and conduction band components of the dressed valence bands, 
and $\xi_{\vec{k}}$ and $E_{\vec{k}}$ are determined by solving 
Eq. \ref{SC}.  We then find the value of $\delta$ consistent with specified 
values of $n_{ph}$ and $\mu$ by observing that 
\begin{equation}
\label{eq:detuning}
\begin{split}
\delta& = \epsilon_{ph} + E_b = \mu - \frac{\partial \epsilon_{mat}}{\partial n_{ph}} + E_{b}\\
&=\mu+E_b+\frac{1}{A}\frac{g}{\sqrt{n_{ph}}}\sum_{\vec{k}}u_{\vec{k}} v_{\vec{k}}.
\end{split} 
\end{equation}
In this way we can solve for all physical quantities as a function of the physical variables 
$\delta$ and $n_{pol}$.  For example in Fig.~\ref{Fig:PhaseDiagram} we plot the chemical potential $\mu$ and 
the photon fraction $n_{ph}/n_{pol}$ as a function of $\delta$ and $n_{pol}$ over the experimentally relevant 
range of these two parameters.  The polariton density is of course not directly controlled experimentally, but  
depends non-linearly on the non-resonant exciton pumping power and on the planar cavity leakage rate in a manner
that can be successfully modeled.

\section{Results} 
\label{results}

\subsection{Exciton-Photon model} 

Thermodynamic properties of the polariton condensate can be predicted on the 
basis of an attractive simplified model that contains only bare exciton and photon degrees of freedom.  
In mean-field theory the ground state condensed exciton ($\Psi_{ex}$) and photon ($\Psi_{ph}$)
fields have identical phases and magnitudes that are determined by minimizing the energy with respect to 
the exciton and photon densities, $n_{ex} = |\Psi_{ex}|^2$ and $n_{ph}= |\Psi_{ph}|^2$. 
In the simplest version of this model no interactions are included.  Because of the photon and 
exciton kinetic energies the ground state condensates are spatially uniform and the energy per unit area is 
\begin{equation} 
\epsilon(n_{ex},n_{ph}) = \epsilon_{ex} n_{ex} + \epsilon_{ph} n_{ph} - 2 \Omega \sqrt{ n_{ex} n_{ph}}.  
\end{equation} 
where $\Omega$, the Rabi coupling, is the matrix element of the matter-photon coupling term in Eq.~\ref{Hamiltonian} 
between the 1-photon/0-exciton and 0-photon/1-exciton states, which we discuss further below.  In the polariton 
condensate the excitons and photons share the same chemical potential:
\begin{eqnarray}  
\label{nonintphex}
\mu = \frac{\partial \epsilon}{\partial n_{ex}} &=& \epsilon_{ex} - \Omega \sqrt{\frac{n_{ph}}{n_{ex}}} \nonumber \\
\mu=  \frac{\partial \epsilon}{\partial n_{ph}} &=& \epsilon_{ph} - \Omega \sqrt{\frac{n_{ex}}{n_{ph}}} .
\end{eqnarray} 
Solving Eq.~\ref{nonintphex} we obtain 
\begin{equation} 
\mu = \epsilon_{LP} =  \frac{\epsilon_{ex}+\epsilon_{ph}}{2} - \sqrt{\Big(\frac{\delta}{2}\Big)^2+ \Omega^2} 
\label{eq:epsilonLP}
\end{equation}
where $\delta = \epsilon_{ph} - \epsilon_{ex}$ is the detuning.  
As expected the chemical potential of a polariton condensate is equal to the energy of a single-polariton
when interactions are neglected.
 
A more realistic version of the exciton-photon 
model can be obtained by adding a term to the energy function to account for the repulsive 
interactions between excitons
\begin{equation} 
\label{energy}
\epsilon = \epsilon_{ex} n_{ex} + \epsilon_{ph} n_{ph} - 2 \Omega \sqrt{ n_{ex} n_{ph}} +\frac{U}{2} n_{ex}^2.
\end{equation} 
where $U$ is the short-range exciton-exciton repulsive interaction.\cite{Tassone1999}
With this change the formula for the exciton chemical potential is modified by replacing the exciton energy by 
a renormalized value containing a mean-field blue-shift: $\epsilon_{ex} \to {\tilde \epsilon}_{ex} = \epsilon_{ex}  + U n_{ex}$.  
The resulting implicit expression for the polariton chemical potential can be reorganized as an expression for the 
chemical potential as a function of polariton density by using the relation
\begin{equation}
\label{eq:nex} 
n_{pol} = n_{ex} + n_{ph} = n_{ex} ( 1 + \frac{\Omega^2}{(\epsilon_{ph}-\mu)^2} ).  
\end{equation} 
It follows that for large positive detuning $n_{pol} \approx n_{ex}$, and $\mu \approx {\tilde{\epsilon}}_{ex}$
which increases strongly with polariton density, whereas for large negative detuning $n_{ex} \approx n_{pol} \Omega^2/\delta^2$, and 
$\mu \approx \epsilon_{ph}$, which is nearly independent of polariton density.
Below we compare our full microscopic results closely with this model of photons 
coupled optically to interacting excitons.  

\subsection{Fermionic Mean-Field Theory}  

The numerical results presented below were obtained by solving the self-consistent-field equations
explained in Section II.  For convenience we consider the case in which the conduction and 
valence band masses are identical, ignore the reduction in two-dimensional 
electron-electron interactions associated with finite quantum well widths, use Bohr radius $a_B^*=\epsilon \hbar^2/(me^2)$ as our length unit
and the excitonic Rydberg $\text{Ry}^*=e^2/(2\epsilon a_B^*)$ as our energy unit. 
For typical GaAs quantum well materials, $m_e=0.067m_0$, $m_h=0.6m_0$, and
$\epsilon=13.18\epsilon_0$\cite{Harrison2005}, yielding $a_B^*\sim115{\AA}$ and
$\text{Ry}^*\sim4.7 meV$.
In our numerical calculation, we choose the band gap as the zero of energy
so that $\epsilon_{ex}=-E_{b} = - 4 Ry^*$, in agreement 
with the narrow well 2D hydrogenic exciton limit.  
In realistic calculations the exciton binding energy is 
substantially reduced by finite well-width effects that allow electrons to spread their charge across the quantum well.
We choose $g=0.5 \text{Ry}^*a_B^*$ for the band-edge photon-induced
interband excitation coupling constant. 
From isolated-polariton calculations, which are equivalent to the dilute-polariton limit of our 
polariton condensate calculations, we find that the relationship between the Rabi coupling and 
the photon-induced transition coupling constant is 
\begin{equation}
\Omega= g \int \frac{d^2 \vec{k}}{(2\pi)^2} \, \phi_{\vec{k}}=\frac{4g}{\sqrt{2\pi}a_B^*},
\end{equation}
where $\phi_{\vec{k}}$ is the momentum-space hydrogenic ground state wave function
in the narrow quantum well limit.  In this way 
we obtain $\Omega  \approx 1 \text{Ry}^*$. 
As we emphasize below, the effective Rabi coupling constant $\Omega$ implied 
by our fermionic mean-field-theory calculations is not constant as it is in
the exciton-photon model. 

\begin{figure}[t]
	\includegraphics[width=1.0\columnwidth]{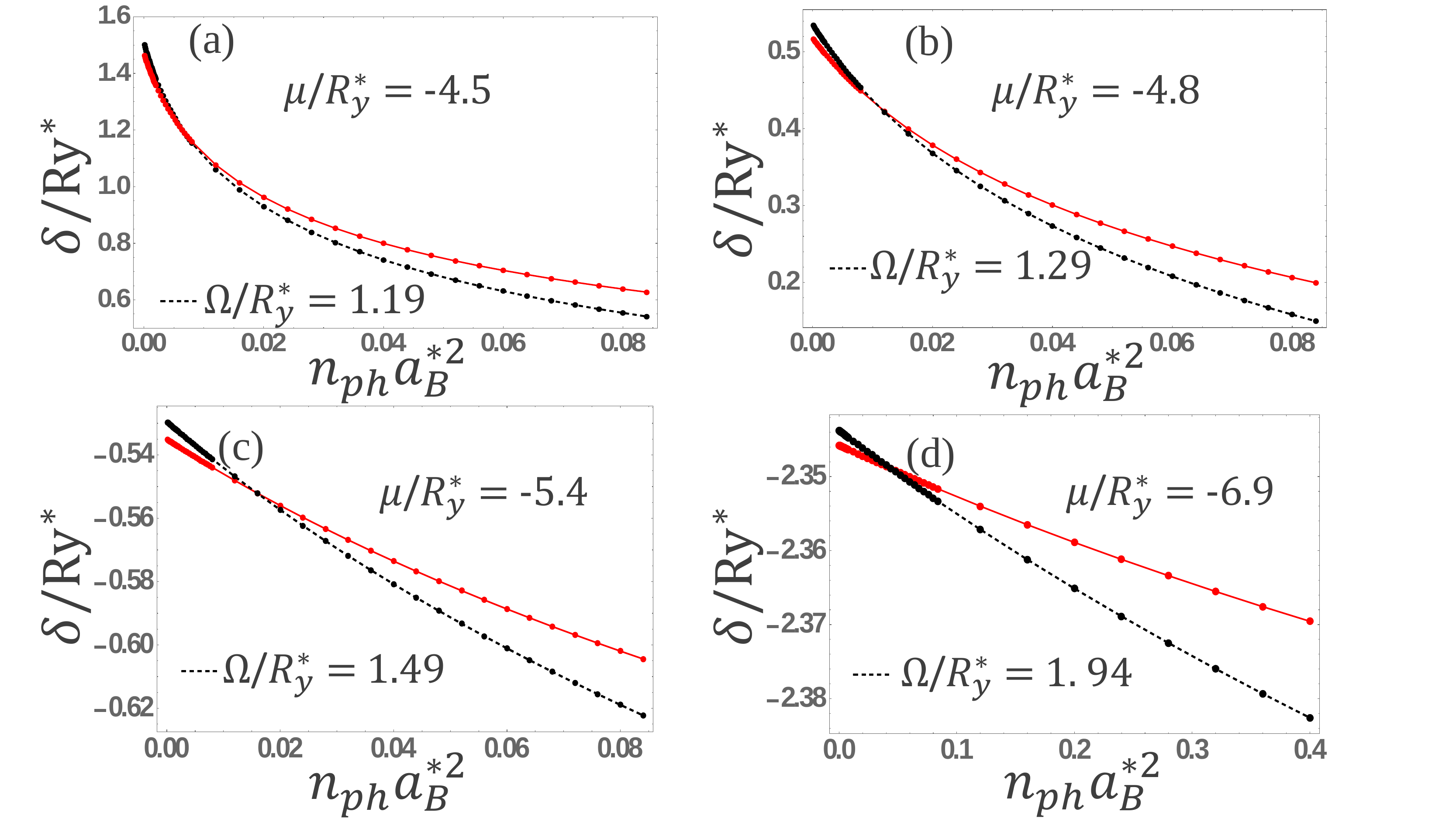}
	\caption{(Color online)
		(a)-(d) Detuning $\delta$ (red-line) at which mutual equilibrium is established 
		as a function of photon density $n_{ph}$ for a  
		series of chemical potentials $\mu$ which lie between the lower-polariton and 
		the isolated exciton energies.  The detuning value predicted by an exciton-photon model 
		is plotted as a black line for comparison. The chemical potential dependent 
		effective Rabi coupling $\Omega$ values listed in the four panels 
		were determined by fitting Eq.~\ref{eq:simple_delta} in the main text to our numerical data.
		}
	\label{Fig:deltavsn}
\end{figure}

We present our results as a function of detuning $\delta$ and polariton density $n_{pol}$. 
The detuning is readily adjusted\cite{Snoke2015} 
experimentally simply by varying the optical excitation location
and using wedged microcavity structures.  
The polariton density can be increased by increasing the intensity of the 
pumping laser used to create a bath of non-equilibrium excitons.
Polariton condensates that are in an effective equilibrium state can however 
be obtained only over a limited range of polariton densities, with a small 
but non-zero threshold.  For very strong pumping, the matter excitations fall out 
of equilibrium with the cavity photons and the pumped steady state 
is that of a standard laser.  Our theory does not address these 
limits on the range of polariton density over which quasi-equilibrium condensates 
can be realized.

The change from $n_{ph}$ to $\delta$ as a control variable
is unique provided that $\delta=f(n_{ph})$ at fixed $\mu$ is invertible, {\it i.e.} that the relationship is 
monotonic.  We establish this property by explicit numerical calculation.  
Fig.~\ref{Fig:deltavsn} demonstrates that detuning $\delta$ is always
a monotonically decreasing function of photon density $n_{ph}$.
We can understand this property by comparing with 
Eq.~(\ref{eq:detuning}), from which we can immediately see
that $\delta$ decreases when $n_{ph}$ increases when we can ignore the
implicit dependence of $u_{\vec{k}}v_{\vec{k}}$ on $n_{ph}$.

In the exciton-photon model calculation corresponding to Fig.~\ref{Fig:deltavsn}, 
we first solve 
\begin{equation} 
\mu = \epsilon_{ex} + U n_{ex}  - \Omega \sqrt{\frac{n_{ph}}{n_{ex}}} 
\end{equation} 
to obtain $n_{ex}$ as a function of $n_{ph}$ and $\mu$ and then use 
\begin{equation}
\label{eq:simple_delta} 
\delta = \epsilon_{ph}+ E_{b} = \mu + E_{b}+ \Omega  \sqrt{\frac{n_{ex}}{n_{ph}}}.
\end{equation}
Comparing Eq.~\ref{eq:simple_delta} with Eq.~\ref{eq:detuning}, the effective $\Omega$ from our
microscopic model is given by:
\begin{equation}
\label{eq:effomega}
\Omega_{eff}=\frac{1}{A}\frac{g}{\sqrt{n_{ex}}}\sum_{\vec{k}}u_{\vec{k}} v_{\vec{k}}.
\end{equation}
In the exciton-photon model, $\Omega$ is a constant whereas in
our microscopic theory its effective value depends on detuning,
as explicitly shown in Eq.~\ref{eq:effomega}. 
In Fig.~\ref{Fig:deltavsn}  the black dashed line is a fit
to the exciton-boson model expression 
for the dependence of detuning on photon density at fixed chemical potential,
and the corresponding values of $\Omega$ are provided in the panel legends.  
The effective values of $\Omega$ obtained in this way characterize 
light-matter interactions and approache the single-polariton value 
when the photon density is small and the photon fraction is small, {\it i.e.} when the 
detuning is positive.  The effective Rabi coupling is expected to be stronger for 
more photon like condensates because the exciton wave function is more spread out in momentum space and 
more localized in real space,\cite{Kamide2010} in agreement with Fig.~\ref{Fig:deltavsn}. 

\begin{figure}[t]
	\includegraphics[width=1.0\columnwidth]{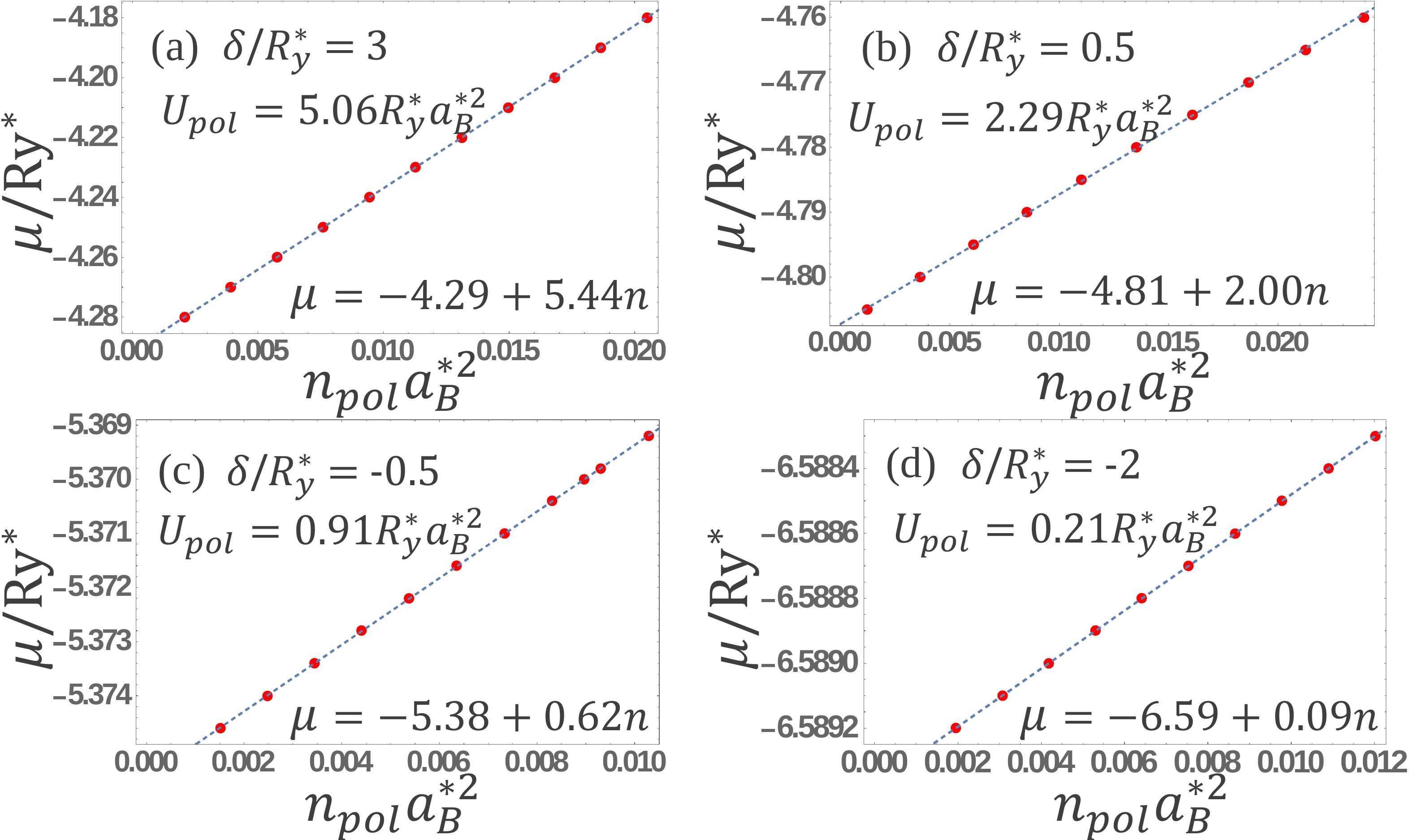}
	\caption{(Color online)
		(a)-(d) Polariton chemical potential $\mu$ as a function of polariton density
		$n_{pol}$ at a series of fixed detuning values.  
	    The dashed line is a linear
		fit to the numerical data from which we determine the 
		effective polariton-polariton interaction strength and lower polariton energy as the slope
		and intercept.(See Eq.(\ref{mu}).) The value of $U_{pol}$ predicted by the exciton-photon model
		(see Eq.~\ref{eq:upol}) is calculated using the Rabi coupling strength defined by the lower polariton energy (Eq.~\ref{eq:rabi}) 
		and U=6 $Ry^*a_B^{*2}$ \onlinecite{Tassone1999}, and is given on the upper left of each panel.  These values should be 
		compared with the microscopic polariton-polariton interactions determined by the
		slopes of the $\mu$ {\it vs.} $n_{pol}$ plots. 
		}
	\label{Fig:muvsn}
\end{figure}

\begin{figure}[t]
	\includegraphics[width=1.0\columnwidth]{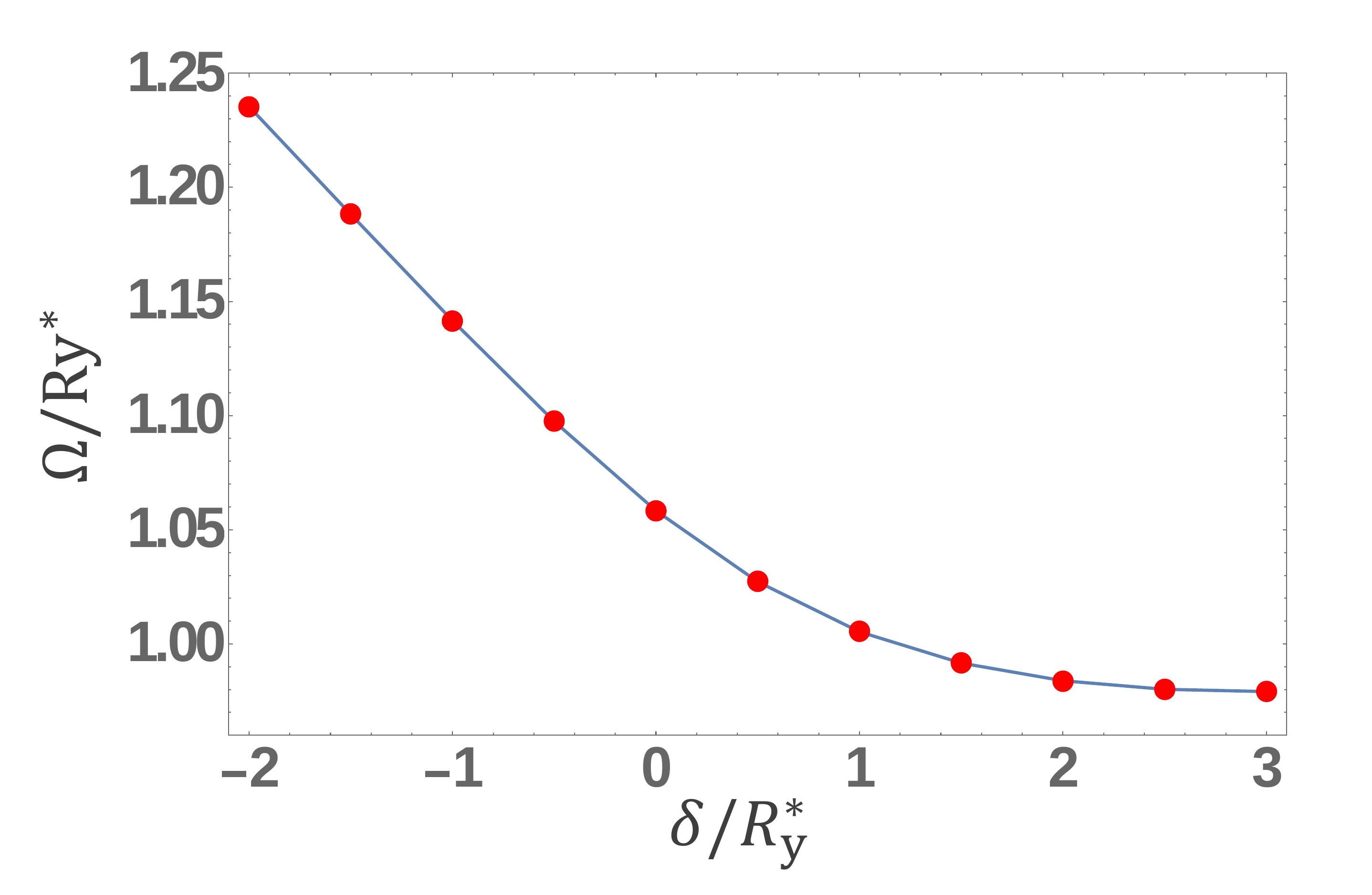}
	\caption{(Color online)
		Effective Rabi coupling $\Omega$ determined by the smallest chemical 
		potential value at which an equilibrium polariton condensate can be formed, as a function of 
		detuning $\delta$.  $\Omega$ is calculated from Eq.~\ref{eq:rabi}.} 
	\label{Fig:rabi}
\end{figure}

As illustrated in Fig. \ref{Fig:muvsn}, we find that for a fixed detuning
there is a minimum value of the chemical potential at which an equilibrium polariton 
condensate can be established, and that the chemical potential increases linearly with
polariton density in the low density limit in
agreement with experiment.\cite{Kasprzak2006, Snoke2015}
We identify the smallest value of the chemical potential at which mutual equilibrium
between photons and matter excitations can be established as the lower-polariton energy,
$\epsilon_{LP}$.  The value of $\epsilon_{LP}$ predicted by the 
microscopic mean-field equations can be compared with the value predicted by the analytic expression  
Eq.~\ref{eq:epsilonLP} by defining another effective Rabi splitting energy $\Omega$ as 
\begin{equation}  
\label{eq:rabi}
\Omega = \frac{\sqrt{(2\epsilon_{ex}+\delta-2\epsilon_{LP})^2-\delta^2}}{2}.
\end{equation}
(Note that $\epsilon_{LP}$ is always smaller than both $\epsilon_{ex}$ and $\epsilon_{ph}$.)  
We find that at the detuning values we have studied  the effective 
$\Omega$ calculated in this way is always close to the bare $1 Ry^*$ value,
as shown in Fig.~\ref{Fig:rabi}. The origin of the stronger Rabi coupling at smaller detuning
is the reduced matter-excitation size in the presence of photons 
discussed above in connection with Fig.~\ref{Fig:deltavsn}.

The initial increase in chemical potential with polariton density 
can be used to define an effective polariton-polariton interaction $U_{pol}$, using 
\begin{equation}
\mu=\epsilon_{LP}+U_{pol} \; n_{pol}.
\label{mu}
\end{equation}
Figs.~\ref{Fig:muvsn}(a)-(d) show that $U_{pol}$ is always 
positive, {\it i.e.} that the polariton-polariton interactions are always repulsive.
These results show that the polariton interaction strength increase monotonically upon going from negative 
to positive detuning as the exciton fraction of the polariton condensate increases.

We can achieve a qualitative understanding of polariton-polariton interaction,
\begin{equation} 
U_{pol} \equiv \frac{\partial \mu}{\partial n_{pol}} \vert_{n_{pol}=0},
\end{equation}
using the simplified exciton-photon model from which we find that 
\begin{eqnarray}
\label{eq:upol}
\frac{U_{pol}}{U}=(\frac{n_{ex}}{n_{pol}})^2_{\vert_{n_{pol}=0}}=\frac{1}{4}(1+\frac{\delta}{\sqrt{\delta^2+4\Omega^2}})^2.
\end{eqnarray} 
The factor on the right side of Eq.~\ref{eq:upol} approaches $1$ at 
strong positive detuning. 
Microscopically the interaction between excitons is repulsive\cite{Keldysh1965,Comte1982} 
and in the dilute limit equal to $6Ry^*a_B^{*2}$.\cite{Tassone1999,Wu&Xue2015} 
Eq.~\ref{eq:upol} accounts for the polariton-polariton interaction that emerges from the matter portion
of the condensate, but not for the fact that the matter excitations are altered by the photon portion of the 
condensate.   Using the effective Rabi coupling defined by Eq.~\ref{eq:rabi}, 
we can compare the prediction of the analytic exciton-photon model expression for U$_{pol}$, 
reported on the upper left of each panel in Fig.~\ref{Fig:muvsn},   
with the values determined by the full microscopic calculations, {\it i.e.} with the slopes of the 
straight-line fits to the $\mu$ {\it vs.} $n_{pol}$ plots.  We see that the polariton-polariton interactions weaken 
even more rapidly as $\delta$ is decreased than in the exciton-photon model.  
This property can be understood in terms of the decrease in exciton size induced by the photon
portion of the condensate mentioned above, which acts to weaken the short range
repulsive exciton-exciton interactions.

\begin{figure}[b]
	\includegraphics[width=1.0\columnwidth]{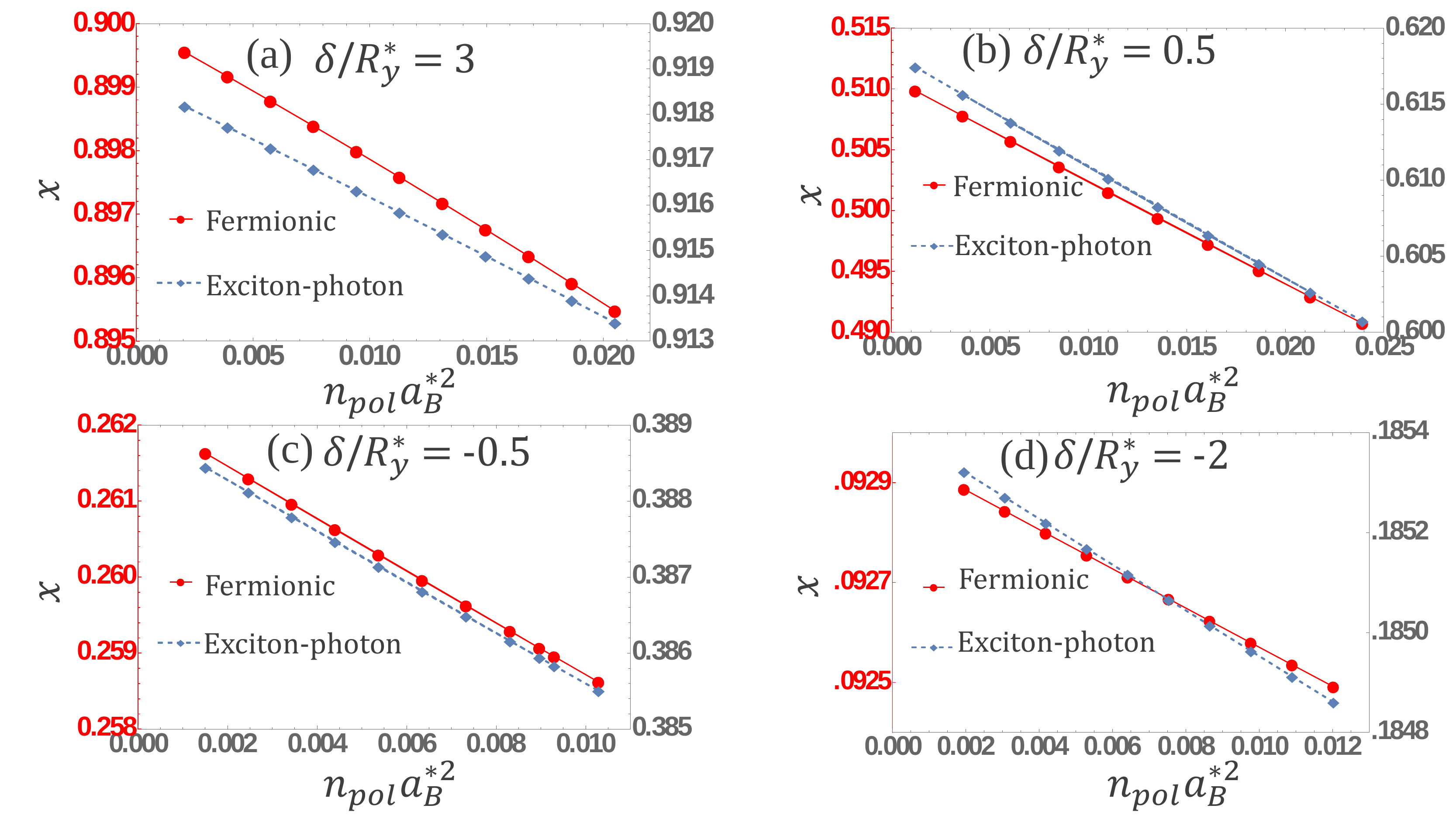}
	\caption{(Color online)
		(a)-(d) Exciton fraction $x=n_{ex}/n_{pol}$ as a function of the density of polaritons 
		$n_{pol}$ at different fixed detuning values $\delta$. The red lines are obtained from 
		microscopic mean-field theory calculations and the blue dashed lines from the analytic expressions
		(Eq.~\ref{eq:nex}) for $x$ in the simplified exciton-photon model.
		Note that red and blue points have different y-values which are shown by red and black marks respectively.
	}
	\label{Fig:exvsn}
\end{figure}

In Figs.~\ref{Fig:exvsn}(a) to (d) we plot the microscopic exciton fraction of the condensate 
$x= n_{ex}/n_{pol}$ as a function of the polariton density at different fixed detuning values and 
compare with the exciton-photon model prediction for the same quantity, Eq.~\ref{eq:nex}.
The simplified model captures the largest trends, namely that polaritons are more exciton-like at more positive
detuning, and that the exciton fraction decreases as the polariton density increases. 
The decrease with polariton density is due to repulsive exciton-exciton interactions which 
increase the effective exciton energy and therefore decrease the effective detuning.  
For example for exciton fractions close to $1$, 
\begin{equation}
x\approx 1-\frac{\Omega^2}{(\epsilon_{ph}-\epsilon_{LP}-U_{pol}n_{pol})^2+\Omega^2}.
\end{equation}
As explained above, polariton-polariton interactions are weaker at smaller values of 
$x$ than predicted by the simplified model.

\subsection{Dressed Bands} 
\begin{figure}[b]
	\includegraphics[width=1.0\columnwidth]{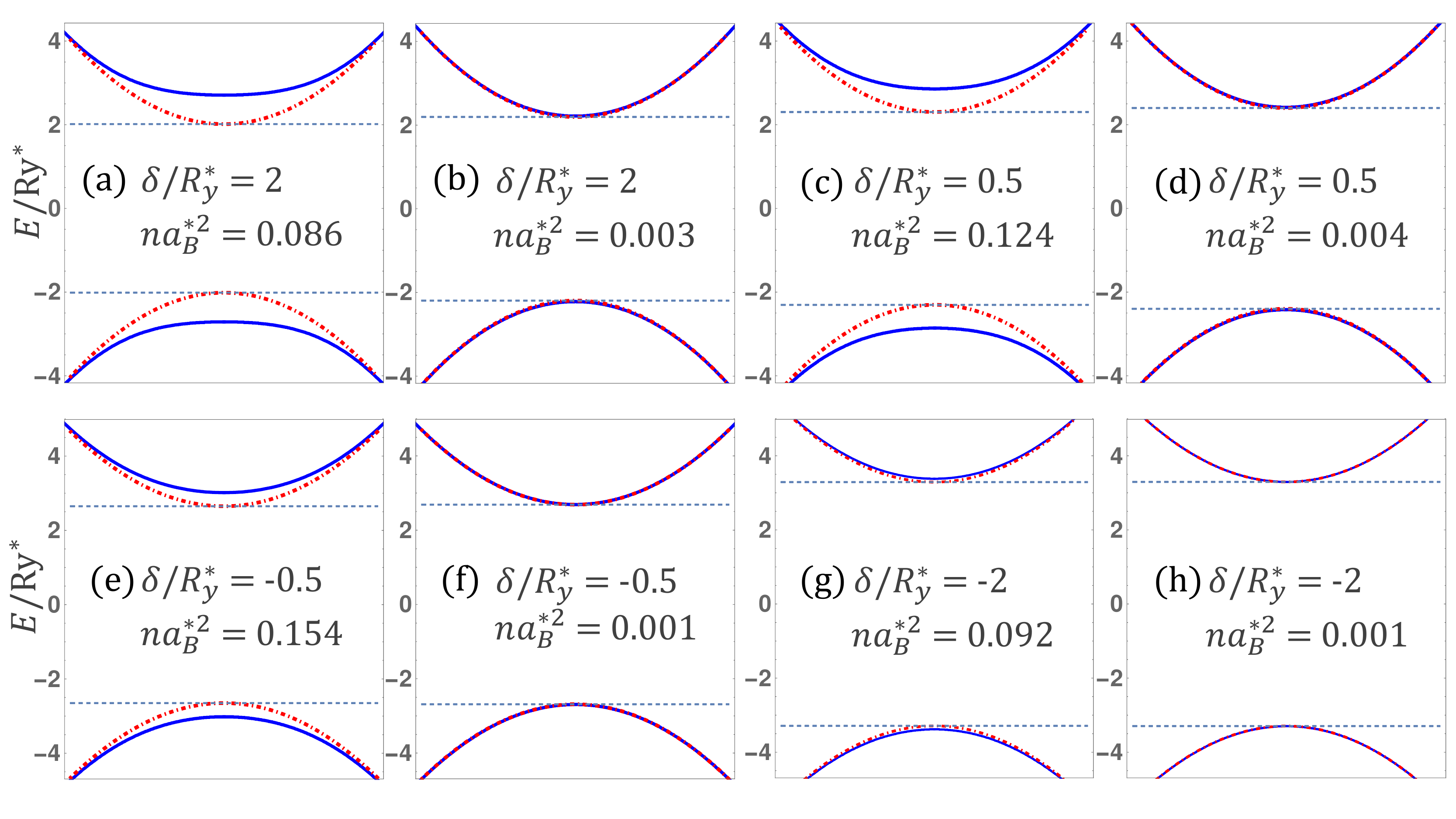}
	\caption{(Color online)
     	(a)-(h) Quasiparticle bands at various detuning and polariton density values.
		The blue lines denote the dressed quasiparticle band structure, while the red dot-dashed lines
		illustrate the bands at the same value of $\mu$ when the self-energies responsible for interband 
		coherence are neglected.  The bare conduction and valence band extrema, marked by dashed horizontal lines, 
		are located at $\vec{k}=0$ in all cases, and have the values $\pm \mu/2$ in the undressed case.  This figure is
		based on calculations with $m_e = m_h$.}
	\label{Fig:quasi}
\end{figure}

The quasiparticle bands of polariton condensates are dressed by both electron-photon and electron-electron interactions.
Results from self-consistent calculations at a series of detuning and polariton density values 
are illustrated in Fig.~\ref{Fig:quasi}.  The property that the dressed bands are coherent combinations of the 
bare conduction and valence bands is the most crucial difference between the steady-state of  polariton condensates and 
the steady state of standard lasers.

\begin{figure}[t]
 \includegraphics[width=1\columnwidth]{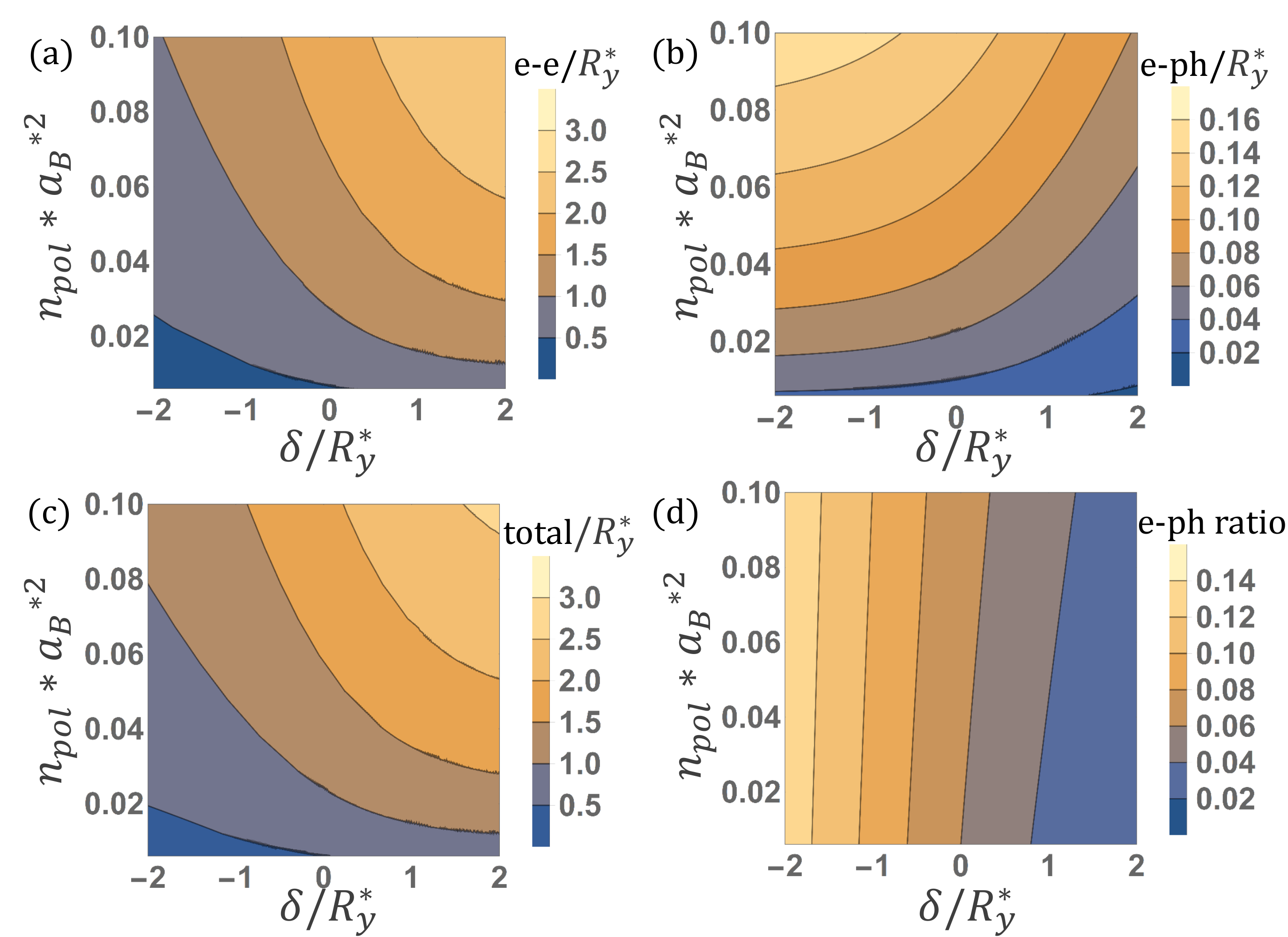}
 	\caption{(Color online)
 		Contour plots of the electron-electron (e-e (a)) and electron-photon (e-ph (b)) contributions to the maximum band-mixing  
		self-energy as a function of polariton density and detuning.  All energies are in $Ry^{*}$ units
		and the polariton density is in $a_{B}^{*-2}$ units.  Panels (c) and (d) plot the total self-energies and the e-ph interaction 
		fractional contribution to the total self-energies.
		Note that the e-e interaction self-energy is largest even when the polariton condensate
		is photon dominated and that the electron-electron interaction contribution is enhanced by the presence of the photon field.}
 		\label{Fig:selfenergy}
 \end{figure}

In the rotating wave picture that we employ the bare bands, plotted as red dot-dashed lines in Fig.~\ref{Fig:quasi}, 
have a gap $\epsilon_{gap} = E_{gap} - \mu$ or simply $- \mu$ because we have chosen the semiconductor band gap
$E_{gap}$ as the zero of excitation energy.  The increase in gap size in the dressed bands, plotted in blue, 
is due to energy level repulsion that is a consequence of mixing between conduction and valence bands.  For negative values of $\mu$,
the BEC limit\cite{Zhu1995} case of interest for polariton condensates, the minimum gap occurs at $\vec{k}=0$ and has the value
\begin{equation} 
\epsilon_{gap} = 2 \sqrt{(\mu/2)^2 + \Delta_{\vec{k}=0}^2},
\end{equation} 
where $\Delta_{\vec{k}}$ has contributions due to both electron-electron interactions and electron-photon interactions,
as specified in Eq.~\ref{SC}.  As noted there the photon contribution to the band mixing self-energy is 
proportional to $\sqrt{n_{ph}}$ and the proportionality constant $g \sim 0.5 \, Ry^* a_{B}^{*}$.  
We can derive a similar expression for the exciton contribution to the 
band dressing self-energy, valid in the low exciton density limit, by examining the linearized gap equation:
\begin{equation}	
\frac{k^2}{2m}\frac{\Delta_k}{2E_{k}}-\frac{1}{A}\sum_{\vec{k}'}V_{\vec{k}-\vec{k}'}\frac{\Delta_{k'}}{2E_{k'}}=\mu\frac{\Delta_k}{2E_{k}},
\end{equation}
and identifying it with the two-dimensional hydrogenic Schrodinger equation.  
We find that 
\begin{equation}
\label{Deltaee}
\begin{split}
&\Delta_{k}=(\frac{k^2}{2m}-\mu)\sqrt{n_{ex}}\phi_{k},\\
&\Delta_{\vec{k}=0}=-\sqrt{2\pi}a_B^* \mu \sqrt{n_{ex}}.
\end{split}
\end{equation}
In Eq.~\ref{Deltaee} $\phi_{k}$ is the 1s hydrogenic wavefunction in momentum space.  
Setting $\mu \to -4 Ry^{*}$, the exciton binding energy, implies a coefficient of 
$\sqrt{n_{ex}}$ that is around $10 Ry^* a_B^*$, more than one order of 
magnitude larger than the coefficient $g=0.5 Ry^*a_B*$
that appears in front of $\sqrt{n_{ph}}$.  The exciton component of the condensate is therefore more 
effective than the photon component in dressing the quasiparticle bands.  
This qualitative point is confirmed by the full microscopic self-consistent calculations
summarized in Fig.~\ref{Fig:selfenergy}.
The band-mixing self-energy plotted in Fig.~\ref{Fig:selfenergy} is the maximum value of $\Delta_{k}$ over values of k.
In most cases, the maximum is located at exactly $\vec{k}=0$ which corresponds to BEC limit discussed above.
Both the electron-electron self-energies and electron-photon self-energies are monotonic 
functions of detuning at fixed polariton density, with the e-e self energies increasing and the electron-photon 
self-energies decreasing with $\delta$.  
The presence of a small photon fraction in the polariton condensate actually increases the 
electron-electron self energy because of the tendency of photons to produce smaller excitons. 
For this reason the electron-electron self energy increases more slowly at fixed polariton density 
than the exciton fraction upon tuning toward positive detuning.

\section{Discussion}  
\label{discussion}

In this paper we have explored a number of properties of equilibrium polariton 
condensates using a microscopic mean-field approximation that becomes exact 
in the limit of low polariton densities.  Quasi-equilibrium steady states of polariton 
condensates are most easily achieved in a polariton condensate when it has a 
substantial exciton fraction, which leads to relatively strong particle-particle 
scattering.  Our microscopic mean-field calculation demonstrates that 
polariton-polariton interactions rates are approximately proportional to the 
square of the exciton fraction, as implied by simplified models that approximate 
the matter portion of the condensate by bare bosonic excitons.    
Indeed polariton condensate formation is closely related to exciton condensate formation. \cite{Keldysh1965,Lozovik1976,Comte1982,Zhu1995,Wu&Xue2015} 
The most important distinction is that, even when a small fraction of the total condensate,
the light portion of the condensate dramatically 
enhances the stiffness of the condensate, promoting longer range 
phase coherence, increasing its robustness in the presence of disorder 
and suppressing the high-exciton-density Mott transition\cite{Liu1998,DePalo2002,Nikolaev2008,Asano2014}
between condensate and incoherent photon-electron-hole plasma states.   

A polariton condensate achieves coherence between matter and light excitations.
The most important consequence of this property is that the mean-field quasiparticle bands of a 
polariton condensate possess coherence between their valence and conduction band components 
driven by both electron-electron and electron-photon interaction self energies.  
We find that the photon contributions to the long-wavelength anomalous self-energy is 
proportional to the square root of the photon density and that the 
matter contribution is, for small densities, also proportional to the square root of the matter 
excitation density. 
However, our calculations show that the coefficients of these dependences are rather 
different and that the electron-electron contribution dominates even when the photon 
fraction of the condensate is relatively large.  The photons provide the glue that holds the 
condensate together because of their large stiffness energy, but the system otherwise
behaves much like a simple exciton condensate.  

We anticipate that the properties of these quasiparticle bands will be important
for future research on the properties of electrically-driven polariton condensates. 
If so, an important issue concerns the coherence strength, which is proportional to the 
ratio of the total band-mixing self-energy to the difference between the energy gap of the 
quantum wells and the chemical potential of the polariton condensate.
Neglecting Rabi splitting, the latter quantity is comparable to the 
exciton binding energy when the polariton density is low.  
Our calculations show that polariton condensates can have substantial interband coherence,
driven mainly by the electron-electron interaction band mixing self-energy.  

The mean-field Hamiltonian of a polariton condensate violates total polariton number conservation.
This property of polariton condensates is analogous to the corresponding properties of  
superconductors and ferromagnets, in which the mean-field Hamiltonians violate 
exact conservation of total particle number and approximate conservation of total spin respectively.
When charge is driven through spatially inhomogeneous superconductors 
and ferromagnets the order parameter condensate is altered because of Cooper pair 
creation or annihilation in the superconductor case and because of spin-transfer torques 
in the ferromagnetic case.  These effects restore the conservation laws.  We anticipate that 
analogous effects will occur when charge is driven through polariton condensates
in which inhomogeneities have been introduced, for example by varying the local detuning, 
to provide convenient electrically tunable polariton sources and sinks.    

\section{Acknowledgment}
This work was supported by ARO Grant No. 26-3508-81,
and by the Welch Foundation under Grant No. 
F1473.

\bibliography{FPC.bib}{}
\bibliographystyle{apsrev4-1}

\end{document}